# Negative differential resistance with graphene channels, interfacing distributed quantum dots in Field-Effect Transistors

Samarth Trivedi[1], and Haim Grebel[2]*

[1]Department of Chemistry, New Jersey Institute of technology (NJIT), Newark, NJ 07102, USA

[2]Electronic Imaging Center and ECE Dept., New Jersey Institute of technology (NJIT), Newark, NJ 07102, USA.  *Email: grebel@njit.edu

**Abstract -** Field effect transistors with channels made of graphene layer(s) were explored.  The graphene layer(s) contacted a distributed array of well-separated semiconductor quantum dots (QDs).  The dots were imbedded in nano-structured hole-array; each filled hole was occupied by one dot.  Differential optical and electrical conductance was observed.  Since Negative Differential Resistance (NDR) is key to high-speed elements, such construction may open the door for new electro-photonic devices.

Graphene [1], a mono, or a few layers of graphite, has attracted a vast interest recently [2,3]. Field effect transistors (FET) demonstrated its unique electrical properties early on [2,4]. Besides monolayer, attention also turned to bi- and multilayered graphene based devices [5]. One may expect unique electrical effects when the mono- or a few-layered graphene is interfaced with semiconductor quantum dots [6-9]. When deposited on porous substrates, such as anodized aluminum oxide (AAO), part of the graphene layer is suspended over the substrate's pores (where the QD are imbedded), enabling study of semi free-standing films [10,11].

The array of pores in the anodized aluminum oxide layer provides us with yet another advantage. Surface plasmon polariton (SPP) are electromagnetic modes, which are confined to an interface between a conductor (graphene in our case) and a dielectric (aluminum oxide on one side of the graphene channel and a protective polymer on the other). The modes decays exponentially away from the surface and, hence concentrate electromagnetic field near the surface (and toward the QDs). The periodic pattern of pores enables coupling of free space radiation to propagating surface modes. If properly designed, the array of pores may facilitate standing surface modes for a strong coupling between electromagnetic radiation and QDs [12-14].

Negative Differential Resistance (NDR) is a nonlinear electronic response. It can be described as a competition between tunneling and thermionic excitations [15]. Resonating effects occur even if the current in the channel is only marginally coupled to the quantum dots, let alone passing through it [16]. Localization of charge [6], gating via the quantum dot ionization [8] and energy transfer [9] while local, may affect the global current-voltage response due to the high charge mobility of the graphene layer. Here we explore the effect of such nonlinear response in graphene channels when interfaced with a distributed array of quantum semiconductor dots.

The schematic of the FET and an SEM picture of the porous substrate are shown in Fig. 1. 150 nm of $SiO_2$ deposited a <100> p-type Si wafer; the Si served as a back gate electrode. 250 nm of Al deposited the $SiO_2$ layer; the Al was later anodized completely per previous recipe [14]. Anodization of the Al resulted in a hole-array with a pitch of ca 100 nm and a hole-diameter of 20-25 nm. The hexagonal hole-array was polycrystalline with a typical domain size of a few microns. Thus, when the pump laser's spot-size is smaller than the domain size, one may couple light to the plasmonic modes more effectively [14]. On the other hand, if the laser spot size covers many domains with slightly varying pitch, broadening of the photoluminescence (PL) signal may be detected (Fig. 2). The CdSe/ZnS QD (Ocean NanoTech, AZ), either with peak luminescence at 590 nm, or at 610 nm were suspended in toluene and drop-casted into the anodized porous substrate. The QDs were coated with octadecylamine to prevent agglomeration while in suspension. The overall size of a coated dot was ca 9 nm and mostly one QD occupied a filled AAO nano-hole (Fig. 1b). Excess dots lying on the substrate surface were washed away. Upon drying, we either deposited the graphene by use of our (inverse) lapping method [11], or by use of transfer of chemical vapor deposited (CVD) graphene [17]. For the latter, we retained the 250 nm thick pmma layer, used for transfer, as a protective coating. The first method yielded a mesoscopic graphene film consisting of 3 to 4 layers, while the second method yielded a more uniform 2-layer film as determined by Raman spectroscopy. A typical domain size of 10-15 microns was similar to both graphene films. The graphene served as the device channel, as well as a semi-transparent layer. However, at these thicknesses, the effective channel might be limited by electrical screening [5]. Photoluminescence data were obtained in confocal arrangement. A 20 mW Ar ion laser beam at wavelength of 514.5 nm was focused to a 10 μm$^2$ spot. The sample was tilted and rotated to produce optimal coupling to surface plasmon polariton (SPP) modes as in [9]. As we shall see below, optimal coupling conditions coincide with normal incidence angle (and hence with normal collection angle due to the confocal configuration).

Fig. 2 shows a typical PL of QD in AAO when the laser beam was focused to a 10 μm$^2$ spot or spanned over 2 mm$^2$. Selective scattering resulted in broadening of the luminescence peak when the interrogation beam covered many domains. Therefore, the PL data was taken with a focused laser beam.

Fig. 3 shows the drain-source current $I_{DS}$ as a function of drain-source voltage, $V_{DS}$ and gate-source voltage, $V_{GS}$, with and without irradiation by 10 mW/cm$^2$ CW, (another) green laser at 532 nm. The graphene channel was deposited by our (inverse) lapping method. The small laser fluence was enough to excite the dots, thus better couple their (positive) charges to the graphene channel. That process left the dot negatively charged, resulting in a clearer NDR effect. Typically, the NDR region (where the current drops to zero) is in the range of $V_{DS}$:[0.3,0.6] V. The composite channel of lapped graphene and QD exhibited p-type characteristics.

Figure 4 shows the electrical and optical data from a transferred CVD grown graphene. A typical valley at $V_{GS}$~0 appears, similarly to the one exhibited by pristine graphene channels [2]. Here, the valley decreased at larger $V_{DS}$ values. The PL followed the electrical trend. The PL as a function of $V_{DS}$ is similar to what has been reported by us in the past for lapped graphene channels [7]: it was enhanced as a function of drain-source potential reaching a maximum and slowly decreasing for larger $V_{DS}$ values.

The mesoscopic channels exhibited p-type behavior indicating that the current was mediated by both graphene and QDs in what may be amounted to a hopping mechanism. In contrast, the transferred CVD grown channels retained its pristine graphene characteristics. In this latter case, better uniformity resulted in current flowing mainly in the graphene layer which implies that channel conductance was controlled by the Dirac points. Nevertheless, a small but clear NDR may be observed at $V_{GS}$~0 and $V_{DS}$~0.3 V. Around that region, $I_{DS}$ depends on $V_{DS}$ linearly (as is the case for most pristine graphene films). Both channels exhibited similar PL trend because of the strong coupling between individual dots and the graphene layer(s). This means that the luminescence was mainly governed by local charge distribution.

Coupling of the pump laser radiation ($\lambda$=514.5 nm in our case) to the channel and collection of the luminescence spectra is best made at an optimal incident and in-plane rotational angles [14]. In our case, a ~100 nm pitch of the hole-array translates to a normal incident angle for the pump (514.5 nm) and normal collection angle for the PL (~600 nm) wavelengths, involving, respectively 6 and 7 hole-symmetry planes. That explains the need of focusing onto a spot-size of no less than 7 μm (7 symmetry planes times a pitch of 100 nm) but no larger than the coherence length of the hole-array domain (~10 μm). The PL data were taken from a spot of ca 10 μm$^2$ compared to an overall device area of 1 cm$^2$. Thus, the global current-voltage curve of our experiments may be understood as the effect of localized charge re-distribution, enabled by the high-mobility graphene layer around the laser spot.

In summary, by taking advantage of semi-suspended graphene channels in FET devices and by coupling the channels to individual quantum semiconductor dots, we demonstrated a novel electro photonic structure. This new class of devices may find possible applications in communication and sensing systems.

Figure Captions

Fig. 1. (a) Schematics of the device configuration. (b) SEM picture of QD-filled AAO. The position of a dot is marked by a yellow circle.

Fig. 2. PL of QD in AAO when the laser beam is focused (a) and de-focused (b). The effect of the cut-off filter is noted at λ=540 nm.

Fig. 3. IDS vs VDS and VGS for (inverse) lapped graphene on AAO: (a) NDR is observed at VGS=-3 V and VDS=0.5 V when illuminated by a 10 mW/cm$^2$ laser at 532 nm and, (b) less pronounced when measured in darkness.

Fig. 4. (a) $I_{DS}$ (in μA) of transferred, CVD grown graphene onto AAO under radiation of 10 mW/cm$^2$ of a 532 nm CW laser. NDR is noted at $V_{GS}$~0 V and $V_{DS}$~0.3 V. The Ar ion laser was focused to a spot in the middle of the channel to obtain, (b) PL (a.u) at 600 nm and (c) PL (a.u) at 580 nm.

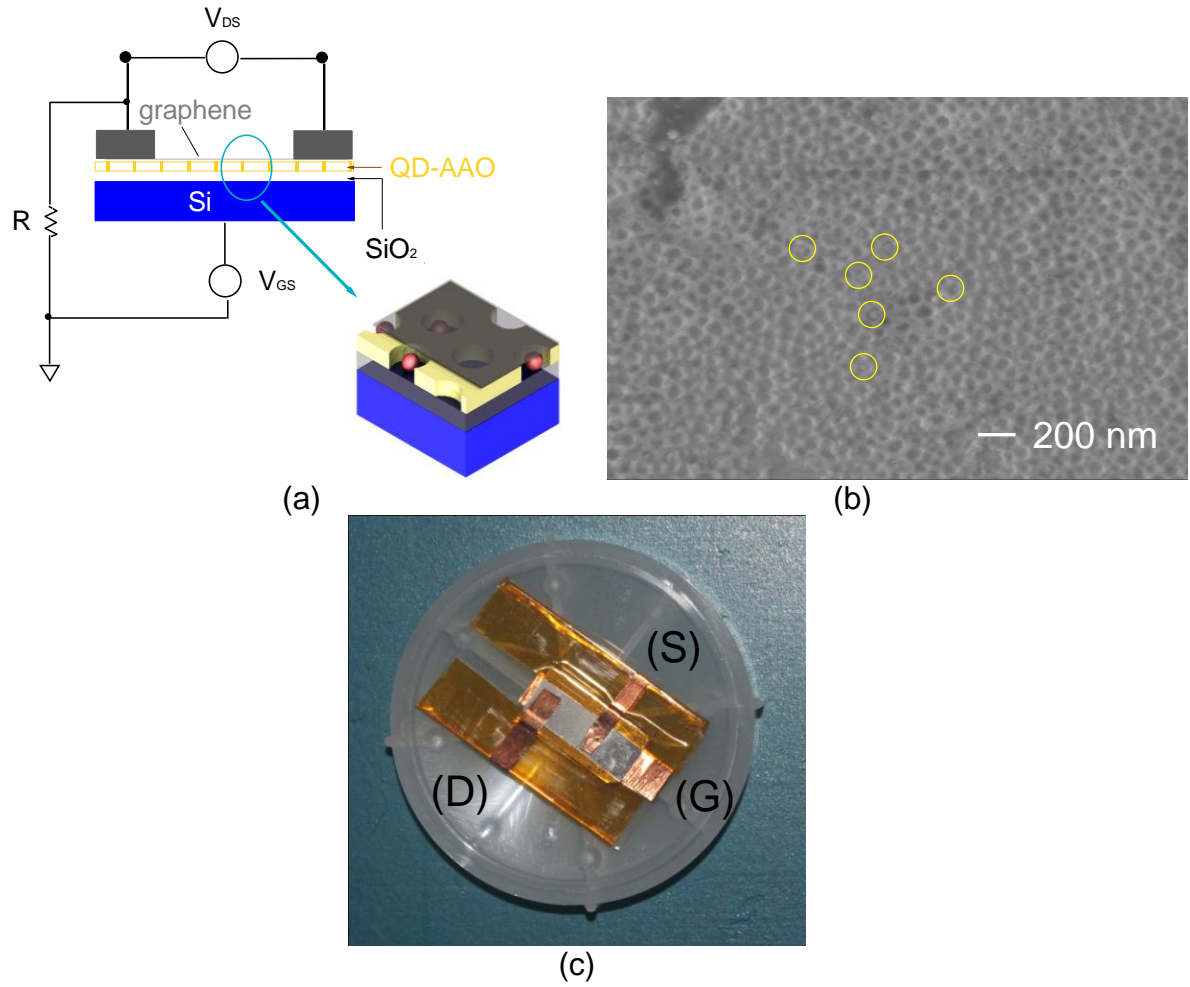

Fig. 1. (a) Schematics of the device configuration. (b) SEM picture of QD-filled AAO. The position of a dot is marked by a yellow circle. (c) 1 cm$^2$ channeled device with a transferred CVD grown graphene: (D), (S) and (G) are Drain, Source and Gate electrodes, respectively. The graphene was depositing the region between the D and S.

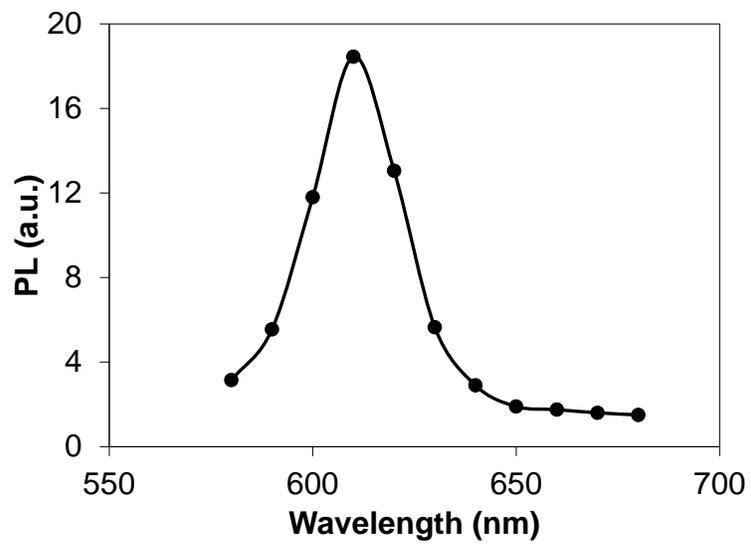

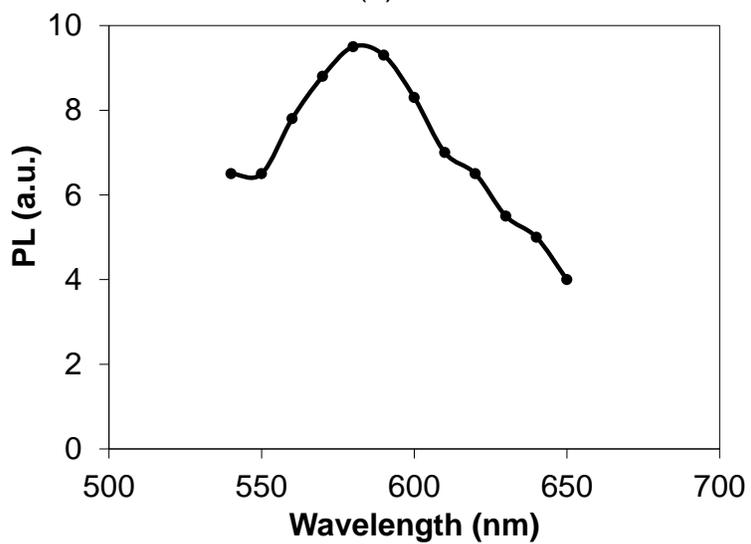

Fig. 2. PL of QD in AAO when the laser beam is focused (a) and de-focused (b). The effect of the cut-off filter is noted at λ=540 nm.

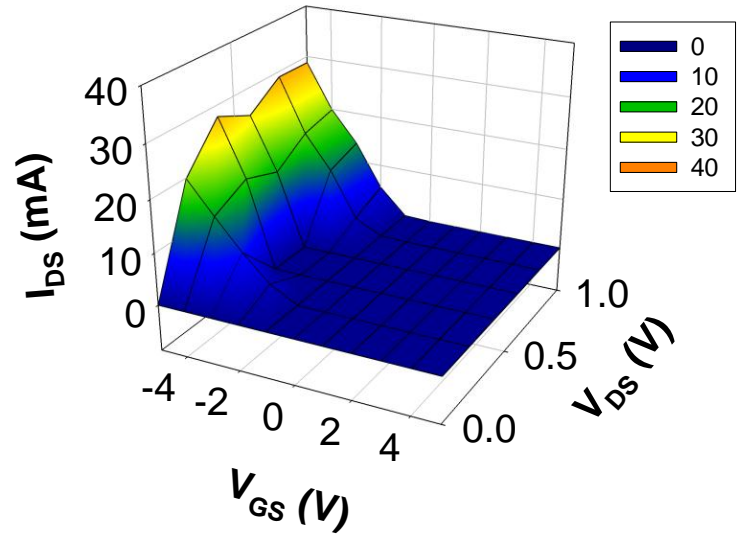

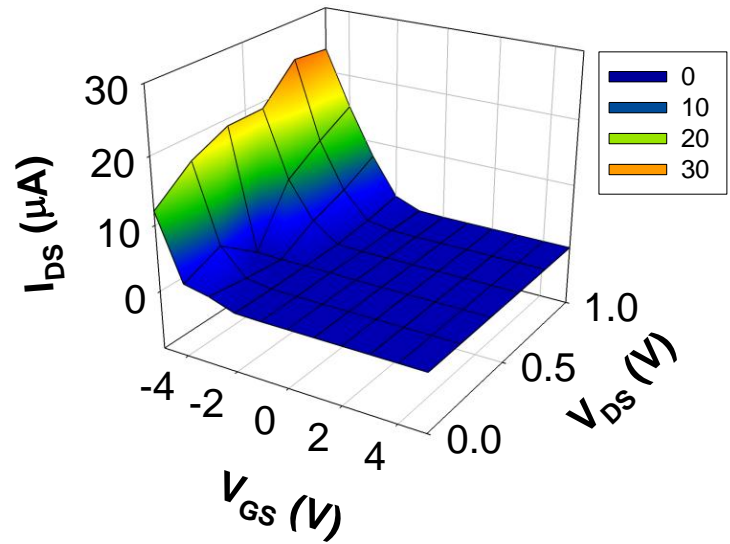

Fig. 2. $I_{DS}$ vs $V_{DS}$ and $V_{GS}$ for (inverse) lapped graphene on AAO: (a) NDR is observed at $V_{GS}$=-3 V and $V_{DS}$=0.5 V when illuminated by a 10 mW/cm$^2$ laser at 532 nm and, (b) less pronounced when measured in darkness.

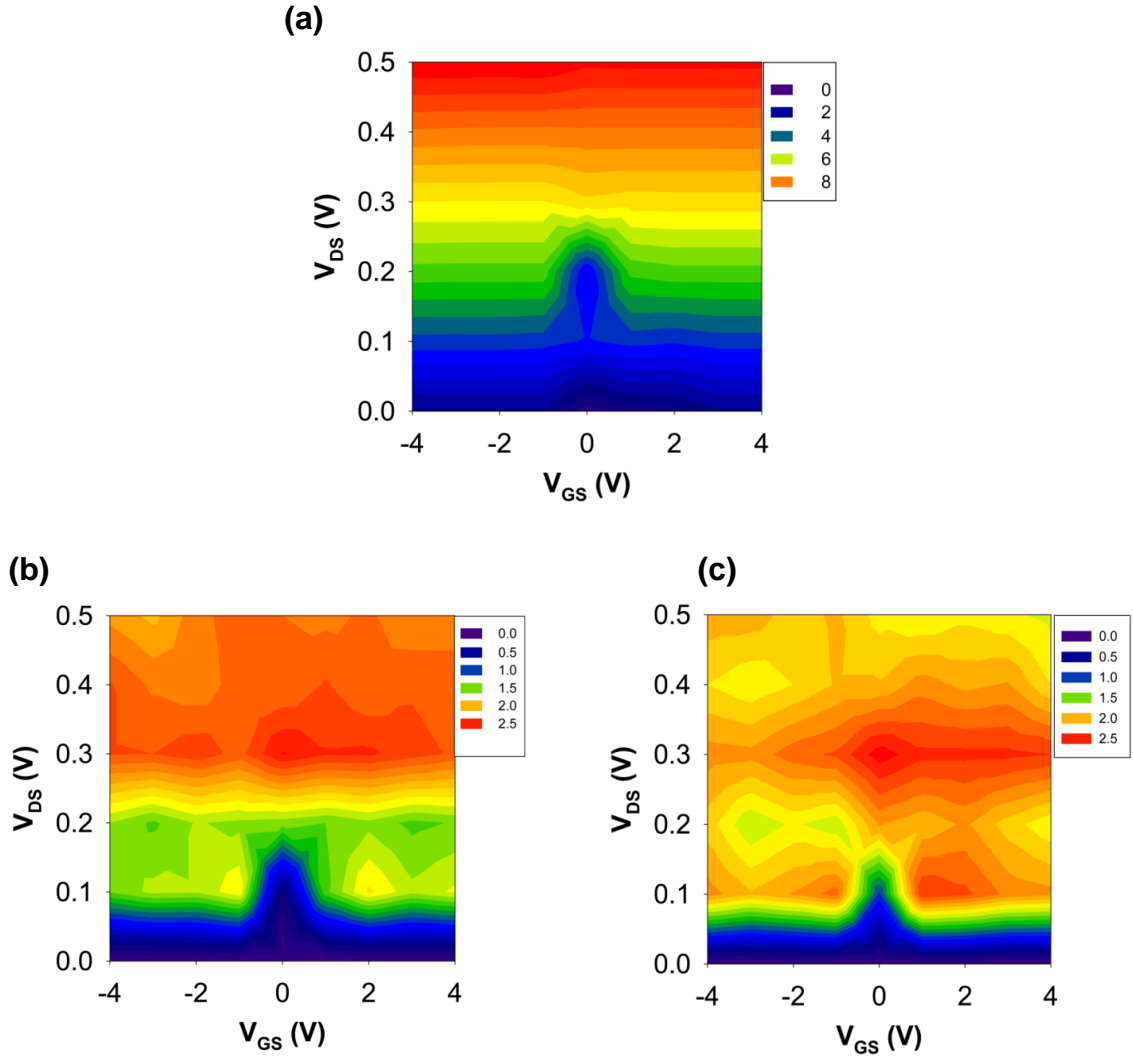

Fig. 4. (a) $I_{DS}$ (in μA) of transferred, CVD grown graphene onto AAO under radiation of 10 mW/cm$^2$ of a 532 nm CW laser. NDR is noted at $V_{GS}$~0 V and $V_{DS}$~0.3 V. The Ar ion laser was focused to a spot in the middle of the channel to obtain, (b) PL (a.u) at 600 nm and (c) PL (a.u) at 580 nm.